\documentclass[twocolumn,superscriptaddress,floatfix,prl]{revtex4-1}
\usepackage{graphicx,amsfonts,amssymb,amsmath,hyperref,hypcap,enumerate}
\usepackage{color}

\newcommand{\ba}{\boldsymbol{a}}

\newcommand{\QQ}{\boldsymbol{Q}}
\newcommand{\RR}{\boldsymbol{R}}
\newcommand{\rr}{\boldsymbol{r}}

\newcommand{\kk}{\boldsymbol{k}}
\newcommand{\dd}{\boldsymbol{d}}

\newcommand{\bb}{\boldsymbol{b}}

\newcommand{\GG}{\boldsymbol{G}}

\begin{document}
\title{Hubbard Model Physics in Transition Metal Dichalcogenide Moir{\' e} Bands}

\author{Fengcheng Wu}
\affiliation{Materials Science Division, Argonne National Laboratory, Argonne, Illinois 60439, USA}

\author{Timothy Lovorn}
\affiliation{Department of Physics, University of Texas at Austin, Austin, Texas 78712, USA}

\author{Emanuel Tutuc}
\affiliation{Department of Electrical and Computer Engineering, Microelectronics Research Center, The University of Texas at Austin, Austin, Texas 78758, USA}

\author{A. H. MacDonald}
\affiliation{Department of Physics, University of Texas at Austin, Austin, Texas 78712, USA}

\date{\today}

\begin{abstract}
Flexible long period moir\' e superlattices form in
two-dimensional van der Waals crystals containing layers
that differ slightly in lattice constant or orientation.
In this Letter we show theoretically that isolated flat moir\' e bands described by generalized
triangular lattice Hubbard models are present in twisted transition metal
dichalcogenide heterobilayers.
The hopping and interaction strength parameters of the Hubbard model can be tuned by varying the twist angle and the three-dimensional dielectric environment.
When the flat moir\'e bands are partially filled, candidate many-body ground states at some special filling factors include spin-liquid states, quantum anomalous Hall insulators and chiral $d$-wave superconductors.
\end{abstract}

\maketitle

\begin{figure}[t]
    \includegraphics[width=1\columnwidth]{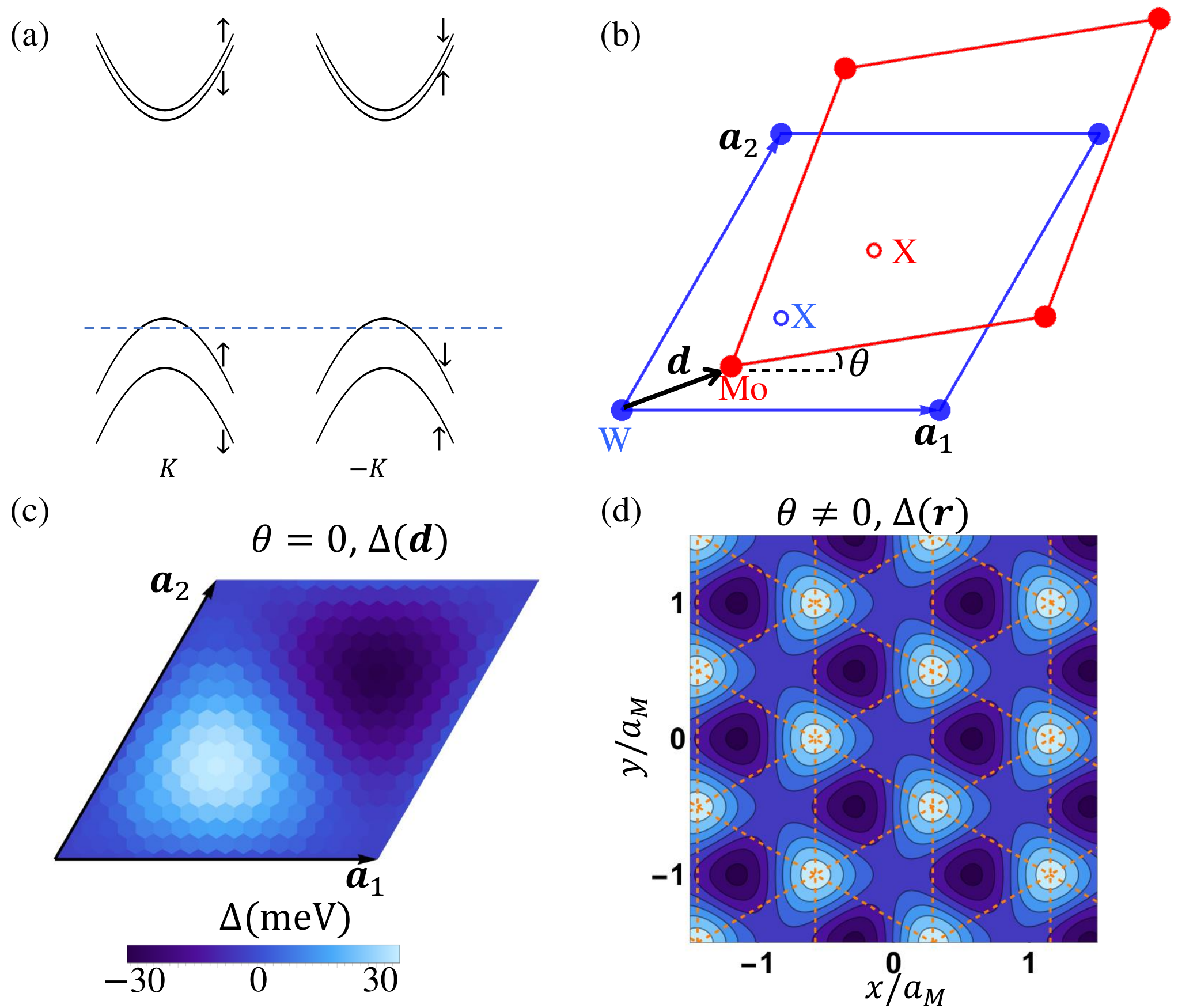}
	\caption{(a) Schematic band structure of monolayer WSe$_2$ with a large (small) spin-splitting
	at valence (conduction) band extrema located at the $\pm K$ valleys.
	(b) AA stacked WX$_2$/MoX$_2$ bilayers with an additional in-plane displacement $\dd$,
	 and a twist angle $\theta$. $\ba_1$ and $\ba_2$ are primitive translation vectors of WX$_2$.
	 (c) Dependence of the WSe$_2$ valence band maximum $\Delta(\dd)$ on displacement $\dd$ in AA stacked
	 WSe$_2$/MoSe$_2$ with zero twist angle.  $\Delta(\dd)$ has triangular lattice periodicity and one maximum per
	 triangular lattice unit cell. (d) When a moir\'e pattern is formed the band maximum variation is magnified from the atomic scale to the moir\'e pattern scale.  The color scales in (c) and (d) are identical and the orange dashed lines in (d) are near-neighbor links that connect $\Delta(\rr)$ maxima.}
	\label{Fig:Moire_Lattice}
\end{figure}

{\it Introduction.---}
Long-period superlattices form when two-dimensional crystals are overlaid with a
small difference in lattice constant or orientation.  When the two-dimensional crystals
are semiconductors or semimetals, their low-energy electronic degrees of freedom
can\cite{Bistritzer2011} be accurately described using continuum models in which commensurability
between the moir{\' e} pattern and the atomic lattice plays no role.  Because the continuum
model Hamiltonians are periodic in space, their single-particle eigenstates satisfy
Bloch's theorem and form bands in momentum space, referred to as moir{\' e} bands.
The moir{\'e} band Hamiltonian acts in a spinor-space whose dimension is
determined by the number of low-energy bands in the host two-dimensional crystal.
In the moir{\' e} band model of twisted bilayer graphene, for example, there are sixteen low-energy bands
corresponding to the bilayer's four triangular sublattices, and to spin and valley.
Mott insulators and superconductivity have recently been discovered in the flat bands of twisted bilayer graphene.\cite{Kim_bilayer, Cao2018Magnetic,Cao2018Super}
In this Letter we construct moir{\' e} band Hamiltonians for holes in twisted
heterobilayers formed from semiconducting transition metal
dichalcogenides (TMDs), which have only two low-energy valence
bands when the chemical potential is within the topmost valence bands as illustrated in Fig.~\ref{Fig:Moire_Lattice}(a), and therefore map to single-band Hubbard models.  We show theoretically
that isolated flat Bloch bands described by generalized triangular lattice Hubbard models
are present in TMD heterobilayers,
and that spin-liquid states are likely to occur when these bands are close to half-filling.
The moir\'e bilayers provide a new solid-state platform to simulate
the Hubbard model, one in which model parameters such as bandwidth, interaction strength,
and band filling are widely tunable. This two-dimensional platform
can be studied at accessible temperatures using a variety of experimental techniques,
for example transport and scanning tunneling microscopy.

To simulate the two-dimensional Hubbard model, it is necessary to identify a
moir{\' e} system in which it is possible to isolate a single band with a two-fold degeneracy.
Because two-dimensional group-VI TMDs such as MoS$_2$ have band extrema
at two inequivalent Brillouin-zone corners \cite{Di2012}, conduction band states occur in groups of four with valley degeneracy and a small spin-splitting due to
spin-orbit interactions.  We therefore focus on the isolation of
moir{\' e} bands formed from orbitals in the valence band, which have a very large spin-splitting [Fig.~\ref{Fig:Moire_Lattice}(a)].  Valley degree of
freedom then faithfully plays the role of spin in Hubbard model.
To avoid any additional degeneracy due to the presence of two layers, we choose to study
heterobilayers.
We consider common-chalcogen TMDs WX$_2$/MoX$_2$ (X= S, Se), which have very similar lattice constants and can have long period moir{\' e} patterns.
Below we focus on the influence
of the moir{\' e} pattern on states near the maximum of the WSe$_2$ valence bands,
which lie inside the MoSe$_2$ gaps and are only weakly coupled to states in MoSe$_2$ due to band offsets.\cite{CZhang2017, wilson2016band}
We note that similar physics can also be realized in TMD bilayers with different chalcogen atoms \cite{Zhang_Shih,Feenstra2018}, for example WSe$_2$/MoS$_2$, which is studied in detail in the Supplemental Material \cite{SM}.

{\it Moir{\' e} potential---}
To derive the valence band moir{\' e} Hamiltonian from first principles
we follow the approach outlined in
Ref.~\onlinecite{Jung2014}, which in the present case requires
an evaluation of the dependence of the WX$_2$ valence band maximum energy $\Delta$
on the relative displacement $\boldsymbol{d}$ between
two layers with identical lattice constants and twist angle $\theta=0$.
The {\it ab initio} calculation was performed
using fully relativistic density-functional theory in
the local-density approximation as implemented in QUANTUM ESPRESSO \cite{giannozzi2009}.
In Fig.~\ref{Fig:Moire_Lattice}(c), we plot numerical values of $\Delta(\dd)$ for the AA stacked
WSe$_2$/MoSe$_2$ bilayer illustrated in Fig.~\ref{Fig:Moire_Lattice}(b).
In the twisted bilayer moir\' e pattern ($\theta \neq 0$), the local value of $\boldsymbol{d}$ changes slowly over the moir{\' e} period ($a_M$) and the valence band maximum,
which serves as a spin-independent external potential, follows the variation of $\boldsymbol{d}$ and varies periodically in space.
Because we are interested only in moir{\' e} periods greatly in excess of the host material
lattice constant ($a_0$), an effective mass approximation
can be used for the band dispersion of the host material.  We choose
$m^* \sim 0.35 m_0$  for  WSe$_2$, where $m_0$ is the free electron mass.
Combining these considerations we obtain the following moir{\' e} band Hamiltonian:
\begin{equation}
\begin{aligned}
&\mathcal{H}=-\frac{\hbar^2\QQ^2}{2m^*}+\Delta(\rr),\\
&\Delta(\rr)=\sum_{\bb}' V(\bb) \exp[i \bb \cdot \rr],
\end{aligned}
\label{Hamiltonian}
\end{equation}
where $-\hbar^2\QQ^2/(2m^*)$ and $\Delta(\rr)$ are the moir\'e band kinetic and potential energies.
The potential $\Delta(\rr)$ shown in Fig.~\ref{Fig:Moire_Lattice}(d) can be accurately
approximated by a Fourier expansion that includes only the six moir\'e reciprocal lattice vectors in the
first shell. Because the potential is real and each TMD monolayer has three-fold-rotational
symmetry, we require  $V(\bb)=V^*(-\bb)$ and $V(\hat{\mathcal{R}}_{2\pi/3} \bb)=V(\bb)$.
Therefore, all six $V(\bb)$ are fixed by $V(\bb_1) = V \exp(i\psi)$, where $\bb_1 = 4\pi/(\sqrt{3} a_M) \hat{x}$.
Fitting to the {\it ab initio} potential energy, we find that $(V, \psi)$ is (6.6meV,$-94^{\circ}$)
for WSe$_2$ on MoSe$_2$ in AA stacking.
The fitting procedure has been described in detail in Refs.~\cite{Wu2017,Wu2018}.
Because  the coupling between the two layers can be modified by external vertical electric field \cite{Tutuc2018} and by pressure \cite{Dean2018}, the moir\'e potential is experimentally tunable.
The unprecedented advantage of van der Waals heterobilayers is that the moir\'e potential
period can be tuned simply by changing the twist angle: $a_M \approx a_0/\theta$.
In the case of WSe$_2$/MoSe$_2$, $a_M$ is about 19 nm at $1^{\circ}$ twist angle.
We note that collective excitations, for example excitons, experience a similar moir\'e potential energy
whose influence has been studied theoretically in Refs.~\cite{Wu2017, Wu2018, Yu2017}.

{\it Hubbard model---}
The length scales relevant to moir{\' e} Hubbard band formation are
the moir\'e period $a_M$ and the spatial extend $a_{W}$
of the Wannier functions associated with the highest-energy moir\'e band,
which is localized around the triangular lattice of moir\'e potential maximum positions.
Near its maximum the moir\'e potential can be approximated by a harmonic oscillator potential: $-\beta V (\delta \rr/a_M)^2/2$, where $\beta= 16 \pi^2 \cos(\psi+120^{\circ})$ for the potential shown in Fig. \ref{Fig:Moire_Lattice}(d).
Within this approximation, $a_W \approx [\hbar^2/(\beta  m^*V)]^{1/4} \sqrt{a_M}$.
Because $a_W/a_M$ scales as $1/\sqrt{a_M}$, we can anticipate that the
highest energy moir{\' e} band flattens with a decrease in the twist angle.


\begin{figure}[t]
    \includegraphics[width=1\columnwidth]{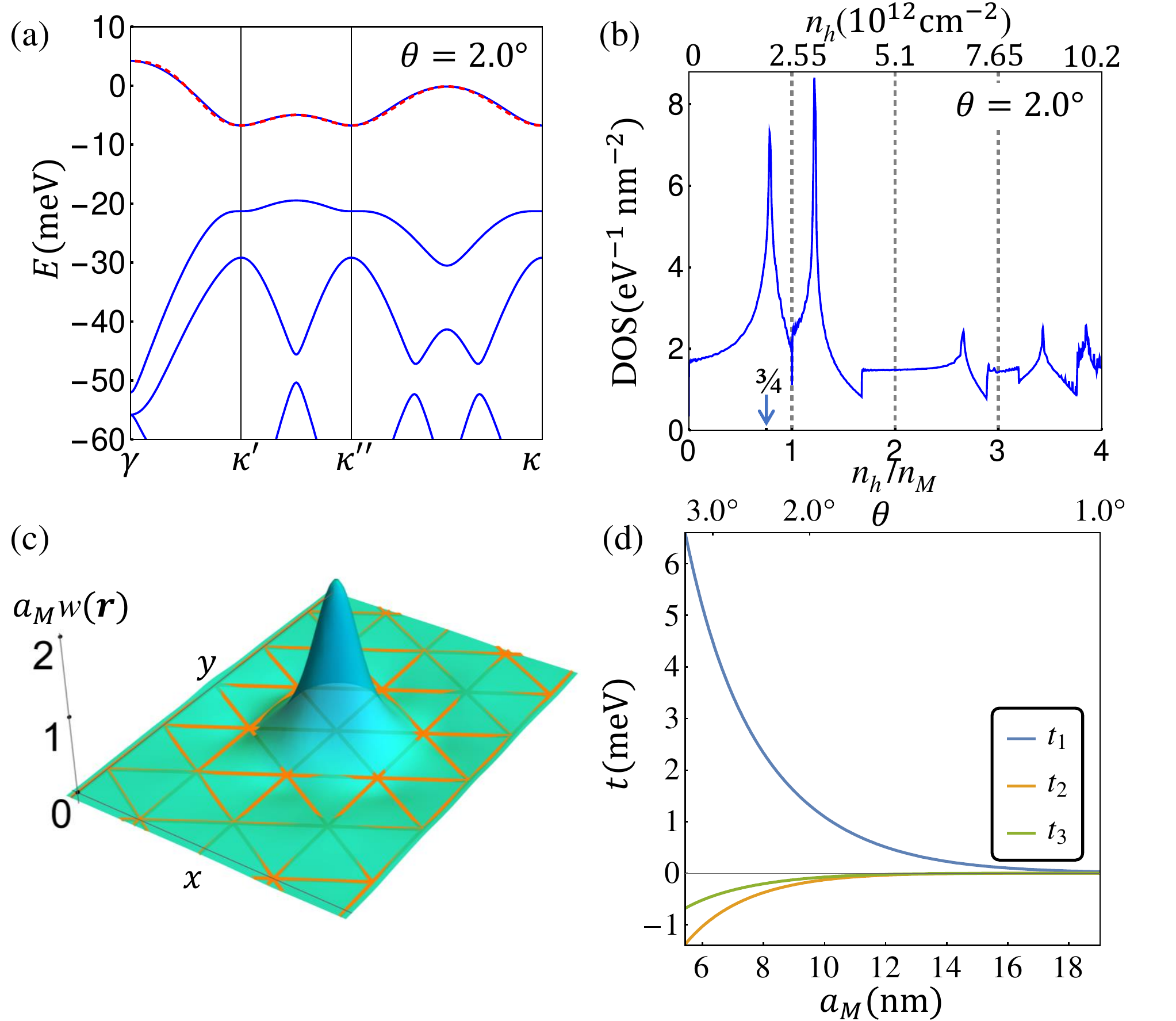}
	\caption{(a) Moir\'e bands at twist angle $\theta=2.0^{\circ}$. The red dashed line is a tight-binding-model fit to the highest valence band that includes hopping up to the third nearest neighbor.
	(b) Density of states as a function of the hole filling factor $n_h/n_M$ (bottom) and the hole density $n_h$ (top).
	(c) The  Wannier function $w(\rr)$ associated with the highest-energy band in (a). $w(\rr)$ is centered on
	 one of the moir\'e potential maxima positions.  The orange lines are the links of the
	 moir\'e triangular lattice.
	 (d) Hopping parameters $t_n$ {\it vs.} neighbor number $n$
	  as a function of moir\'e period $a_M$ (bottom) and twist angle $\theta$ (top).}
	\label{Fig:Band_Structure}
\end{figure}

In Fig.~\ref{Fig:Band_Structure}(a) we plot the moir{\' e} bands of
WSe$_2$ on MoSe$_2$ obtained by diagonalizing the moir{\' e} Bloch
Hamiltonian $\mathcal{H}(\kk)$ in a plane wave representation:
\begin{equation}
\langle \kk+\boldsymbol{g}' | \mathcal{H} | \kk+\boldsymbol{g} \rangle =-\delta_{\boldsymbol{g}',\boldsymbol{g}} \frac{\hbar^2 |\kk+\boldsymbol{g}|^2}{2m^*}
+ V(\boldsymbol{g} '-\boldsymbol{g}),
\end{equation}
where $\boldsymbol{g}$ and $\boldsymbol{g}'$ are moir\'e reciprocal lattice vectors.
The highest valence moir{\' e} band at $\theta=2.0^{\circ}$ is separated from other bands by an energy gap and has a narrow bandwidth ($\sim$11 meV). This isolated flat band can be described by a tight-binding model on a triangular lattice:
\begin{equation}
H_0 = \sum_{\tau=\uparrow,\downarrow}\sum_{\RR, \RR'} t(\RR'-\RR) c_{\RR\tau}^\dagger c_{\RR' \tau},
\label{H0}
\end{equation}
where $\RR$ represents the triangular lattice formed by the moir\'e potential maximum positions, and $\tau$
is a valley index.
In Fig.~\ref{Fig:Band_Structure}(b), we show the density of states (DOS) of the single-particle moir\'e bands as
a function of hole density, which is strongly enhanced by the moir\'e potential, and has sharp peaks at
moir\' e band saddle points.
The flat band energy dispersion can be accurately fit
by including hopping up to the third nearest neighbor.
Figure~\ref{Fig:Band_Structure}(d) shows the hopping parameters $t_n$
as a function of moir\'e period $a_M$, where $t_n$ connects the $n$th nearest neighbors.
The hopping parameters are real, $|t_1|$ is dominant over $|t_{2, 3}|$,
and all three hopping parameters decrease exponentially with increasing $a_M$.

\begin{figure}[t]
    \includegraphics[width=1\columnwidth]{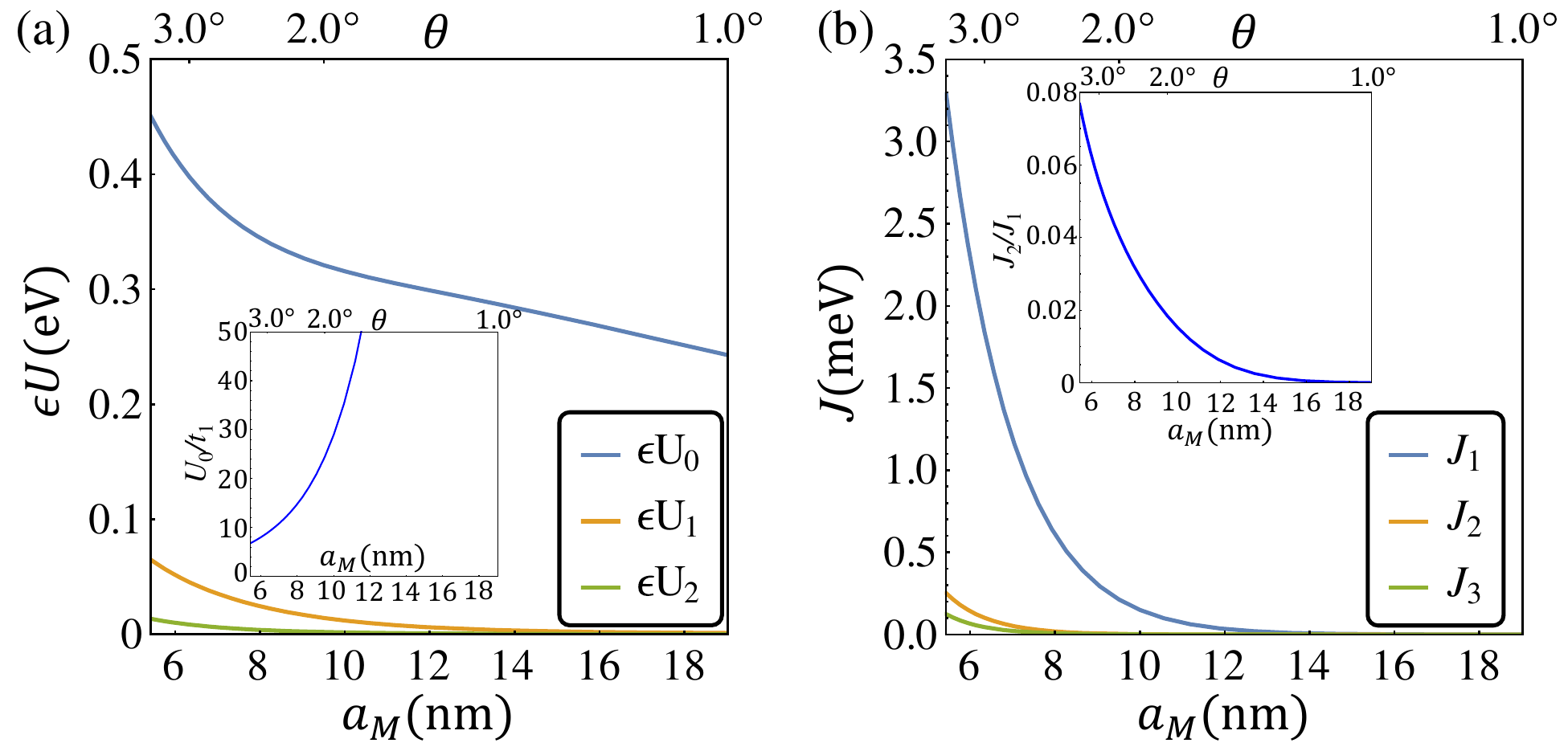}
	\caption{(a) Hubbard model repulsive interaction parameters $\epsilon U_n$,
	 and (b) spin exchange interactions $J_n$ as a function of moir\'e period $a_M$ and twist angle $\theta$.
	 The insets of (a) and (b) respectively show the ratios $U_0/t_1$ and $J_2/J_1$.
	 The semiconductor background dielectric constant $\epsilon$ was set to 10
	 for the evaluation of $U_n$.  Smaller effective values of $\epsilon$ are applicable when the
	 dielectric environment is engineered to maximize interaction strength. For example, $\epsilon$ is about 5 when hexagonal boron nitride is used as the dielectric layer.\cite{Dean2010}  }
	\label{Fig:Interaction}
\end{figure}

Figure~\ref{Fig:Band_Structure}(c) plots the Wannier wave function $w(\rr)$ constructed
from the isolated band's Bloch states.
The spatial extent $a_W$ of this localized wave function increases
with moir{\' e} pattern period, in agreement with the estimate above,
but its ratio to $a_M$ decreases.
Correspondingly the on-site Coulomb repulsion energy
$U_0 \sim e^2/( \epsilon a_W)$ decreases slowly as the moir\'e period increases.
It follows that the ratio of $U_0$ to the bandwidth increases very quickly with $a_M$, and that
that electronic states formed when the moir\'e band is partially occupied by electrons become strongly correlated.
The effective dielectric constant $\epsilon$ in the bilayer is sensitive to the three-dimensional dielectric
environment out to vertical distances $\sim a_M$ from the bilayer, allowing the
strength of correlations at a given orientation angle to be adjusted over a wide range.
To simulate a Hubbard model with short-range repulsion, we assume that
a metallic screening layer is close to the TMD bilayer, but separated from it
by a dielectric. Such a metallic layer, formed by graphene for example,
could also act as a gate that controls the filling factor of the moir\'e band.
In a simple image-charge approximation,
the electron-electron interaction potential is $\tilde{U}(\rr)=(e^2/\epsilon)[r^{-1}-(r^2+D^2)^{-1/2}]$, where $D/2$ is the vertical distance between the metallic layer and the TMD bilayer.
When $\tilde{U}(\rr)$ is projected onto the isolated band Wannier states, and the negligible
overlap between Wannier orbitals centered on different sites is noted,
the interaction Hamiltonian reduces to the generalized Hubbard form:
\begin{equation}
H_1=\frac{1}{2}\sum_{\tau,\tau'}\sum_{\RR,\RR'} U(\RR'-\RR) \, c_{\RR,\tau}^\dagger c_{\RR',\tau'}^\dagger c_{\RR',\tau'} c_{\RR,\tau}.
\label{H1}
\end{equation}
In Fig.~\ref{Fig:Interaction} we plot values of repulsive interaction $U_0$ (on-site), $U_1$ (nearest-neighbor) and $U_2$ (second-nearest-neighbor) as a function of moir\'e period $a_M$.  These results were calculated
using $D=3$ nm, and $U_{1,2}$ are therefore strongly suppressed compared to $U_0$.
As expected from the scaling analysis above, $U_0$ decreases only slowly as $a_M$ increases.

Equation~(\ref{H0}) combined with (\ref{H1}) describes a generalized Hubbard model on a triangular lattice.
The isolated band is fully occupied when the TMD bilayer is charge neutral.
By inducing hole carriers, the band becomes partially occupied.
When the isolated band is completely emptied by hole doping, the carrier density is $n_M=2/(\sqrt{3} a_M^2/2)$, where the factor of 2 accounts for the Hubbard model spin degeneracy.
We find that $n_M = 0.64 \times 10^{12} (\theta[^\circ])^2$ cm$^{-2}$, implying that the full range of band fillings is accessible by electrical gating
for $\theta$ less than $\sim 4^\circ$.
In the following we discuss possible moir\' e band
ground states at $1/2$ and $3/4$ hole doping.

{\it Half filling---}
When the isolated band is half filled, there is one electron per moir\'e unit cell.
As illustrated in Fig.~\ref{Fig:Interaction}(a), $U_0 \gg t_1$ is satisfied even for a relatively large dielectric constant
employed to obtain these estimates.
The strong onsite repulsion suppresses double occupation of moir\'e lattice sites
and gives rise to a Mott insulator ground state with only spin degrees of freedom
at low energies.
In the large $U_0$ limit, the Hubbard model can be mapped to the spin Heisenberg model:
\begin{equation}
H_S = \sum_{\RR, \RR'}' J(\RR'-\RR) \boldsymbol{S}_{\RR}\cdot \boldsymbol{S}_{\RR'},
\end{equation}
where $\boldsymbol{S}$ is the $S=1/2$ spin operator,
$J$ is a spin exchange coupling energy,
and the prime on the sum indicates that each pair of sites is
counted only once.
Using $t/U$ perturbation theory \cite{MacDonald_tU} to calculate the exchange interactions
up to the third nearest neighbors, we find that: $J_1 = 4(t_1^2/U_0)[1-7(t_1/U_0)^2]$, $J_2 =
 4t_2^2/U_0+4t_1^4/U_0^3$ and $J_3=4t_3^2/U_0+4t_1^4/U_0^3$.
Here we have expanded to second order in $t_{2, 3}$, but to
fourth order in $t_1$ because $|t_1|\gg |t_{2,3}|$.
The numerical values of $J_n$ are plotted in Fig.~\ref{Fig:Interaction}(b).

The properties of triangular-lattice Heisenberg models have been thoroughly
investigated in previous work.
When only nearest neighbor coupling $J_1$ is non-zero, the ground state
has three-sublattice 120$^{\circ}$ long range antiferromagnetic order.
This antiferromagnetic state becomes unstable when the second nearest neighbor coupling $J_2$ exceeds
a critical value.
For the quantum spin-1/2  Heisenberg model on triangular lattice, a spin liquid phase has been
found in the parameter region $0.06 \lesssim J_2/J_1 \lesssim 0.17$.\cite{White2015, Sheng2015}.
As shown in Fig.~\ref{Fig:Interaction}(b), $J_2/J_1$ exceeds 0.06 when the twist
angle $\theta$ is larger than 3.0$^\circ$, which makes the spin liquid state likely to occur.
In our case, $J_3$ is also non-zero but its small magnitude seems unlikely to significantly
alter earlier estimates of phase boundaries.
The arrival of moir\'e band  strong correlation physics
motivates new studies of Heisenberg models with exchange coupling to further neighbors.

\begin{figure}[t]
    \includegraphics[width=1\columnwidth]{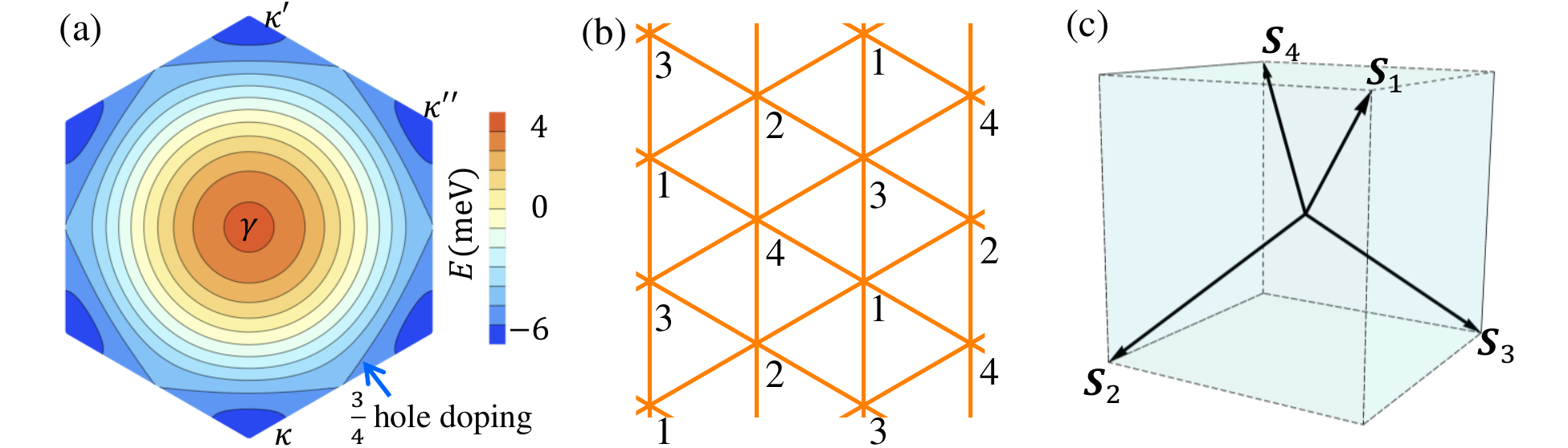}
	\caption{(a) Energy contours in the moir\'e Brillouin zone for the highest energy band in Fig.~\ref{Fig:Band_Structure}(a). (b) Magnetic order on a triangular lattice with four-sublattice tetrahedral antiferromagnetic order. The magnetic moment directions on the four sublattices are  specified by the corresponding arrows in (c).}
	\label{Fig:magnetic}
\end{figure}

{\it $3/4$ filling---} At 3/4 hole doping, flat bands with only nearest-neighbor hopping  have a
van Hove singularity and the corresponding Fermi surfaces are perfectly nested.
As shown in Figs.~\ref{Fig:Band_Structure}(b) and \ref{Fig:magnetic}(a), these features are largely retained in realistic models because remote neighbor hopping is weak.
The nesting vectors are $\bb/2$, where $\bb$ is a first-shell moir\'e reciprocal lattice vector.
One candidate ground state has the four-sublattice tetrahedral antiferromagnetic order\cite{Martin2008} illustrated in Fig.~\ref{Fig:magnetic}.
This magnetic order fully gaps the nested Fermi surface,
and gives rise to a quantum anomalous Hall insulator\cite{Martin2008} with quantized Hall conductivity of $e^2/h$ at
$T=0$. The tetrahedral order is non-coplanar,
and results in a scalar spin chirality: $\chi= \boldsymbol{S}_{i} \cdot (\boldsymbol{S}_{j} \times \boldsymbol{S}_{k})$. Thermal fluctuations at finite temperature will destroy long-range magnetic order in two dimension. However, the
chirality $\chi$ is an Ising order parameter which
can persist even at finite temperatures and support an anomalous Hall effect.
Another candidate state has four-sublattice collinear antiferromagnetic order with site-dependent spin moments.\cite{Nandkishore2012} This state has gapless charge excitations at the Fermi energy in one spin component only, and therefore is a half metal that supports spin currents. In close competition with these magnetic states, there is also an instability towards chiral $d$-wave superconductivity from repulsive interactions.\cite{Nandkishore2012SC, Nandkishore2014} In the renormalization group analysis,
$d$-wave superconductivity has been found to be the leading weak coupling instability at $3/4$ filling.\cite{Nandkishore2014}

{\it Discussion---}
For twist angles smaller than around 3.5$^{\circ}$, the highest energy
WSe$_2$ valence moir{\' e} band provides a realization of the triangular lattice
Hubbard models.  For the special case of half-filling the system provides
a realization of quantum spin models on triangular lattices.  Although the
triangular lattice is frustrated,  the spin-model ground state is a relatively conventional
anti-ferromagnet when only nearest neighbor interactions are present.
Our calculations demonstrate that twist angles can be turned to regimes in which
spin-liquid states are expected.  The estimated spin-interaction energy scales
are on the meV energy scale, making the low-temperature properties of
these quantum spin systems accessible at dilution fridge temperature scales.
The competition between strongly correlated states in the moir\'e band Hubbard model can be tuned by the twist angle, the dielectric environment, and by strain that generates anisotropy for the triangular lattice.
Furthermore, moir{\' e} band Hubbard model realizations also allow
strongly correlated electron systems to be studied in new ways.  For example
by examining how the carrier density depends on gate voltages it is possible to
extract the Hubbard model chemical potential as a function of carrier density,
and in this way to quantitatively extract among other properties, the size of charge
gaps expected at $1/2$ filling, and in some cases
also at $3/4$ filling.

One of the most interesting possibilities offered by TMD moir{\' e} band
systems is that of measuring spin transport characteristics in strongly
correlated electron systems and comparing them directly with charge transport characteristics.
Single layer TMD systems can be optically driven\cite{mak2012control,mak2014valley,Hao2016} into steady
states with valley (and therefore spin) dependent chemical potentials,
allowing them to be used as spin and charge reservoirs,
and as spin-polarization detectors.
These capabilities allow for measurements of coupled spin and
charge transport in strongly correlated electron systems, a topic of
great theoretical interest\cite{Huse2012,Karrasch2014},
and an important goal of cold-atom Hubbard model simulation
efforts\cite{Esslinger2010,Hart2015,Nichols2018}.

FW thanks I. Martin for valuable discussions.
Work at Austin was supported by the Department of Energy, Office of Basic Energy Sciences
under contract DE-FG02-ER45118 and  award \# DE-SC0012670, and by the Welch foundation under grant TBF1473.
Work at Argonne National Laboratory was supported by the Department of Energy, Office of Science,
Materials Science and Engineering Division.
The authors acknowledge HPC resources provided by the Texas Advanced Computing Center (TACC) at The University of Texas at Austin.

\bibliographystyle{apsrev4-1}
\bibliography{refs}

\clearpage
\begin{center}
\textbf{Supplemental Material}
\end{center}

\section{Moir\'e bands in $\text{WSe}_2$/$\text{MoS}_2$ bilayers}
In the different-chalcogen heterobilayer WSe$_2$/MoS$_2$, the top-most valence bands are also primarily associated with WSe$_2$ layer.
There is a relatively large difference in lattice constants  between WSe$_2$ and MoS$_2$, and the mismatch $\delta$ is $|a_0-a_0'|/a_0 \approx 3.9\%$, where $a_0 \approx 3.32 $ \AA ~ and $a_0'\approx 3.19 $ \AA ~ are respectively lattice constants of WSe$_2$ and MoS$_2$. Due to the mismatch $\delta$, there is a moir\'e pattern even in rotationally aligned WSe$_2$/MoS$_2$ bilayer. Experimentally, scanning tunneling microscopy measurement has identified a spatial variation with an amplitude of 50 meV in the valence band maximum energy of aligned WSe$_2$/MoS$_2$ bilayer in AA stacking.\cite{Zhang_Shih} We study moir\'e bands in this system, and the analysis is similar to that presented in the main text. The moir\'e potential that acts on the valence band states of WSe$_2$ is still approximated by:
\begin{equation}
\Delta(\rr)=\sum_{\bb}' V(\bb) \exp[i \bb \cdot \rr]
\end{equation}
where the summation over $\bb$ is again restricted to the six moir\'e reciprocal lattice vectors in the first shell, which are given by: $\bb_j \approx \theta \GG_j \times \hat{z} -\delta \GG_j$. Here $\theta$ is the rotation angle, $\delta$ is the lattice constant mismatch, and $\GG_j$ is the reciprocal lattice vectors of monolayer WSe$_2$ in the first shell. We also require that $V(\hat{\mathcal{R}}_{2\pi/3} \bb)=V(\bb)$ and $V(\bb)=V^*(-\bb)$, and all six $V(\bb)$ are therefore fixed by $V(\bb_1) = V \exp(i\psi)$. By fitting to the experimentally measured moir\'e potential \cite{Zhang_Shih}, we obtain that $(V, \psi)$ is (5.1 meV,$-71^{\circ}$) for WSe$_2$ on MoS$_2$ in AA stacking. The moir\'e period is determined by $a_M \approx a_0/\sqrt{\theta^2 +\delta^2}$, which is about  8.5 nm in the aligned case ($\theta=0^{\circ}$) and decreases as the twist angle increases.

\begin{figure}[t!]
    \includegraphics[width=1\columnwidth]{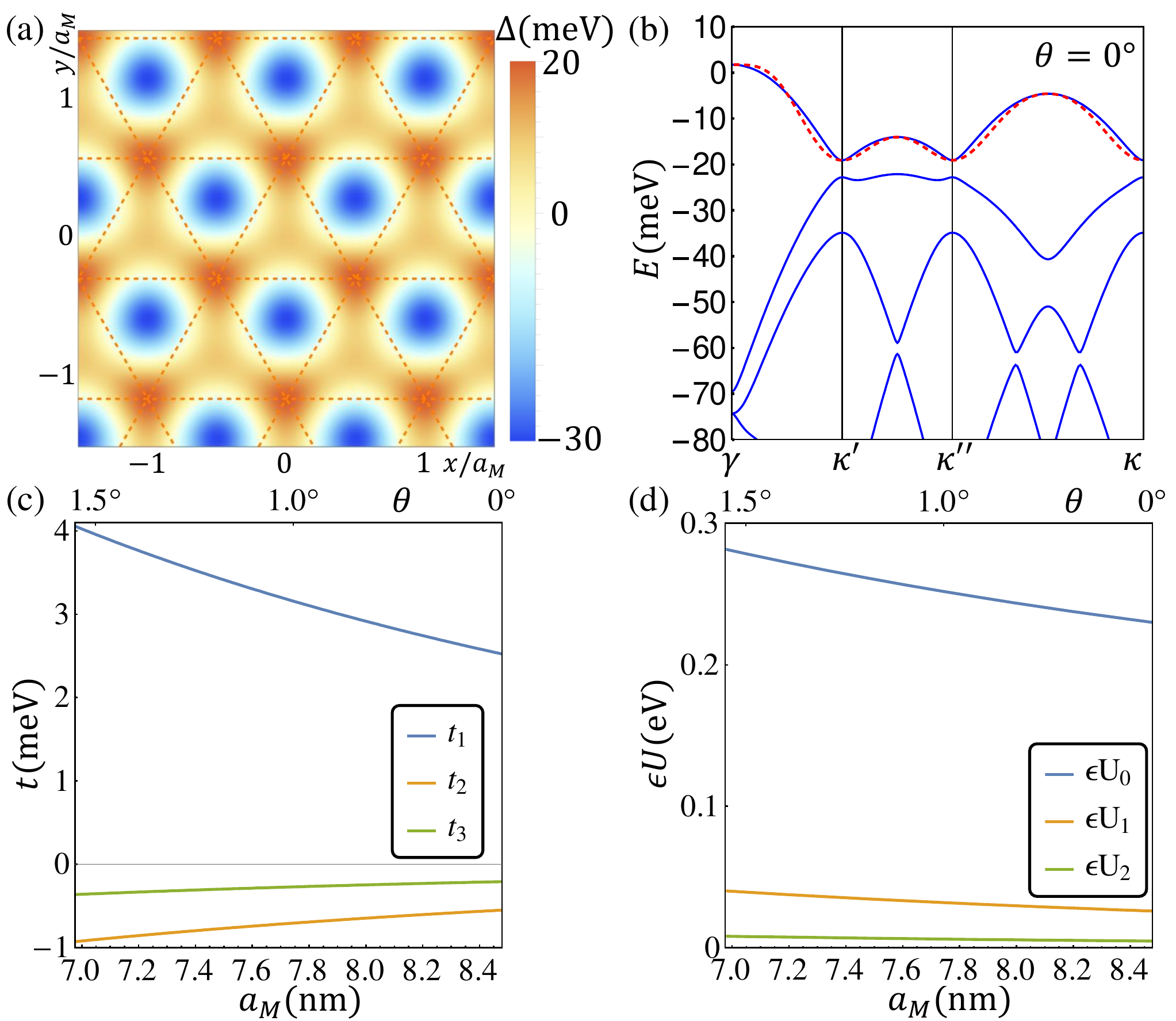}
	\caption{(a) Moir\'e potential energy and (b) moir\'e bands for valence states of WSe$_2$ aligned with MoS$_2$ in AA stacking.
           (c) Hopping parameters $t_n$ and (d) Hubbard model repulsive interaction parameters $\epsilon U_n$
	  as a function of moir\'e period $a_M$ (bottom) and twist angle $\theta$ (top).}
	\label{Fig:WSe2MoS2}
\end{figure}

Results for this system  are summarized in Fig.~\ref{Fig:WSe2MoS2}. At $\theta=0^{\circ}$, the highest valence moir{\' e} band is isolated from other bands and has a bandwidth about 20 meV. The Hubbard repulsive interaction can be at least an order of magnitude larger compared to the hopping parameters. Therefore, Hubbard model physics can also be realized in WSe$_2$/MoS$_2$ bilayer, and the candidate many-body ground states discussed in the main text are applicable as well.

\section{Discussion on Moir\'e potential}
Density-functional theory (DFT) with local or semilocal exchange and correlation functionals fails to capture the long-range van der Waals (vdW) interactions between layers in vdW structures, and leads to inaccurate estimation of binding energies.\cite{Nieminen2012} DFT with random-phase approximation (RPA) correction is required to account for the nonlocal vdW interactions; however, RPA correction is computationally expensive and is beyond the scope of this work.\cite{Nieminen2012} Our primary interest is on the position dependence of the valence band maximum energy, which varies in the moir\'e pattern mainly due to the change in the {\it local} atomic coordination between the two layers. For this reason, we expect that the moir\'e potential will depend less sensitively on the non-local vdW interaction, and DFT with local-density approximation (LDA) can provide a qualitative estimation of the moir\'e potential. As shown in the main text, LDA predicts a moir\'e potential with an amplitude about 60 meV in AA-stacked WSe$_2$/MoSe$_2$ bilayer, which is comparable to the experimentally measured moir\'e potential in a WSe$_2$/MoS$_2$ bilayer.\cite{Zhang_Shih}. This provides a strong justification for our proposal that uses TMD bilayers to simulate Hubbard model physics.

We have focused on the AA stacked bilayer. In TMD heterostructures with a
small difference in lattice constant or orientation, there is another distinct stacking configuration denoted as AB.\cite{Wu2018} AA and AB stacking are distinguished by a 180$^{\circ}$ rotation of the top layer. We find that moir\'e potential is typically shallower in AB stacking compared to AA. In the case of AB stacked WSe$_2$/MoSe$_2$, LDA calculation leads to a moir\'e potential with an amplitude about 10 meV for valence band states associated with WSe$_2$. For this system, the highest valence moir{\' e} band is energetically isolated from other bands when the twist angle $\theta$ is less than $1.5^{\circ}$, and therefore, Hubbard model physics can be realized at small twist angles. In homobilayers like WSe$_2$/WSe$_2$, the layer degree-of-freedom can be effectively removed by applying an external vertical electric field. Thus, spin-1/2 Hubbard model can also be studied in twisted homobilayers.

We have used a rigid rotation picture for the moir\'e pattern, assuming that there is no lattice relaxation. Corrections to this rigid-rotation model will become important only when the moir\'e period is much larger compared to the width of grain boundaries.

\end{document}